\documentclass[aps,prd,preprint,superscriptaddress]{revtex4-1}
\usepackage{color}
\usepackage{amsmath}
\usepackage{graphicx}
\usepackage{latexsym}
\usepackage{ulem}
\usepackage{subfig}

\newcommand{\bfk}{{\mathbf k}}

\newcommand{\hatn}{\hat{n}}
\newcommand{\wtj}[6]{\left(\begin{array}{@{}c@{\; }c@{\; }c@{}}#1 & #2 & #3\\[-5pt]#4 & #5 & #6\end{array}\right)}
\newcommand{\wnj}[9]{\left\{\begin{array}{@{}c@{\ }c@{\ }c@{}}#1 & #2 & #3\\[-1pt]#4 & #5 & #6\\[-1pt]#7 & #8 & #9\end{array}\right\}}

\begin{document}

\preprint{RESCEU-5/17}

\title{CMB B-mode auto-bispectrum \\ produced by primordial gravitational waves}

\author{Hiroaki W. H. Tahara}
\affiliation{Research Center for the Early Universe (RESCEU), Graduate School of Science,
The University of Tokyo, Tokyo 113-0033, Japan}
\affiliation{Department of Physics, Graduate School of Science,
The University of Tokyo, Tokyo 113-0033, Japan}

\author{Jun'ichi Yokoyama}
\affiliation{Research Center for the Early Universe (RESCEU), Graduate School of Science,
The University of Tokyo, Tokyo 113-0033, Japan}
\affiliation{Department of Physics, Graduate School of Science,
The University of Tokyo, Tokyo 113-0033, Japan}
\affiliation{Kavli Institute for the Physics and Mathematics of the Universe (Kavli IPMU), WPI, UTIAS,
The University of Tokyo, Kashiwa, Chiba 277-8568, Japan}

\date{\today}

\begin{abstract}
Gravitational waves from inflation induce polarization patterns in the cosmic microwave background (CMB).
It is known that there are only two types of non-Gaussianities of the gravitaional waves in the most general scalar field theories having second-order field equations.
One originates from the inherent non-Gaussianity in general relativity, and the other from a derivative coupling between the Einstein tensor and a kinetic term of the scalar field.
We calculate polarization bispectra induced by these non-Gaussianities by transforming them into separable forms by virtue of the Laplace transformation. It is shown that future experiments can detect only the new one if the latter coupling parameter takes an extremely large value, which, however, does not cotradict the current observational data.
\end{abstract}

\maketitle

\section{Introduction}
Inflation in the early Universe \cite{Starobinsky:1980te,Guth:1980zm,Sato:1980yn,Linde:1981mu,Linde:1983gd} 
 produces both primordial density perturbations \cite{Mukhanov:1981xt,Guth:1982ec,Starobinsky:1982ee,Hawking:1982cz}
and gravitational waves \cite{Starobinsky:1979ty,Rubakov:1982df}
out of quantum fluctuations.
(For a review of inflation, see e.g. \cite{Sato:2015dga}.)
The simplest model of slow-roll inflation predicts that they are almost Gaussian and that the primordial density perturbations have only small non-Gaussianities of the order of the slow-roll parameters \cite{Acquaviva:2002ud,Maldacena:2002vr}.
Deviations from the simplest model may result in non-Gaussianities large enough to be detected through temperature and E-mode polarization bispectra of the cosmic microwave background (CMB), and CMB experiments \cite{Bennett:2012zja,Ade:2015ava} targeted such large non-Gaussianities to verify the model of inflation. 
The non-Gaussianities are classified to several types which have distinct wavenumber dependence
mostly for the sake of simplicity of calculations
(e.g. \cite{Gangui:1993tt,Verde:1999ij,Wang:1999vf,Komatsu:2001rj,Maldacena:2002vr,Babich:2004gb,Chen:2006nt,Senatore:2009gt}).
However, up to now even the most pricise CMB observation has not detected any statistically
significant deviation from Gaussianity yet \cite{Ade:2015ava}.

On the other hand, primordial non-Gaussianities of the gravitational waves differ from those of the curvature perturbations. 
They are known to be classified to only two types \cite{Gao:2011vs} in the framework of the generalized G-inflation \cite{Kobayashi:2011nu} which is originally the most general single-field inflation model based on the Horndeski theory \cite{Horndeski:1974wa,Deffayet:2011gz} with second-order field equations.
One is inherent in the general relativity and always appears in the primordial non-Gaussianities of the order of unity.
The other is originated from a certain kind of a non-minimal derivative coupling of the scalar field with the Einstein tensor.
Detection of the former would provide an evidence that the standard quantum field theory calculation in curved spacetime works well, whereas that of the latter would be a smoking gun of modification of gravity.

Since primordial density perturbations cannot give rise to B-mode polarization \cite{Seljak:1996gy,Zaldarriaga:1996xe,Kamionkowski:1996zd,Kamionkowski:1996ks}, its primary auto-bispectrum just reflects the non-Gaussianities of gravitational waves.
In this sense, the B-mode auto-bispectrum is suitable for investigating the non-Gaussianities of tensor perturbations and worthy of consideration. So far only the power spectrum of B-mode polarization has been detected which is induced by not primordial gravitational waves but lensing of E-modes and Galactic dust emission \cite{Hanson:2013hsb,Ade:2014afa,Naess:2014wtr,Ade:2014xna,Ade:2015tva}.
Thus there has been little study concerning the B-mode auto-bispectrum with a few exception of models with parity violation \cite{Shiraishi:2013vha,Shiraishi:2013kxa}.

The rest of the paper is organized as follows.
We first state our notation of bispectrum in Section \ref{notation}. 
In Section \ref{horndeski} we show the non-Gaussianity of tensor perturbations in the generalized G-inflation with the Lagrangian of the Horndeski theory.
In Section \ref{calculation} we calculate the bispectrum under linear evolution. 
Finally we show the results in Section \ref{results} and conclude in Section \ref{conclusions}.

\section{CMB bispectrum and reduced bispectrum\label{notation}}
We decompose a CMB anisotopy map $A(\hatn)$ into a set of coefficients of spherical harmonics.
\begin{equation}
A(\hatn)=\sum_{\ell m} a_{\ell m} Y_{\ell m}(\hatn).
\end{equation}
The CMB bispectrum is the three-point correlation of the $a_{\ell m}$'s.
\begin{equation}
B^{m_1 m_2 m_3}_{\ell_1 \ell_2 \ell_3} = \langle a_{\ell_1 m_1}a_{\ell_2 m_2}a_{\ell_3 m_3} \rangle.
\end{equation}
If the Universe is statistically isotropic, then the bispectrum can be angle averaged without any loss of information, that is,
\begin{equation}
B_{\ell_1 \ell_2 \ell_3} = \sum_{m_1 m_2 m_3} \wtj{\ell_1}{\ell_2}{\ell_3}{m_1}{m_2}{m_3} B^{m_1 m_2 m_3}_{\ell_1 \ell_2 \ell_3}, 
\end{equation}
where $\wtj{\ell_1}{\ell_2}{\ell_3}{m_1}{m_2}{m_3}$ is the Wigner-$3j$ symbol.
We also define the reduced bispectrum \cite{Komatsu:2001rj} as
\begin{equation}
b_{\ell_1 \ell_2 \ell_3}=G_{\ell_1 \ell_2 \ell_3}^{-1}B_{\ell_1 \ell_2 \ell_3},
\end{equation}
with
\begin{equation}
G_{\ell_1 \ell_2 \ell_3} \equiv
\mathcal{I}_{\ell_1 \ell_2 \ell_3}^{0~ 0~ 0},
\end{equation}
where we define
\begin{equation}
\mathcal{I}_{\ell_1 \ell_2 \ell_3}^{s_1 s_2 s_3} \equiv
\sqrt{\frac{(2\ell_1+1)(2\ell_2+1)(2\ell_3+1)}{4\pi}}\wtj{\ell_1}{\ell_2}{\ell_3}{s_1}{s_2}{s_3}.
\end{equation}
However $G_{\ell_1 \ell_2 \ell_3}$ vanishes when $\ell_1+\ell_2+\ell_3=\mathrm{odd}$, and in such a case we redifine $G_{\ell_1 \ell_2 \ell_3}$ by using identities of Wigner-$3j$ symbols as \cite{Kamionkowski:2010rb}
\begin{equation}
G_{\ell_1 \ell_2 \ell_3} \equiv
\frac{2\sqrt{\ell_2(\ell_2+1)\ell_3(\ell_3+1)}}{[\ell_1(\ell_1+1)-\ell_2(\ell_2+1)-\ell_3(\ell_3+1)]}
\mathcal{I}_{\ell_1 \ell_2 \ell_3}^{0-1~ 1}.
\end{equation}

\section{\!\!\!\!Primordial non-Gaussianities in the generalized G-inflation\label{horndeski}}
The generalized G-inflation \cite{Kobayashi:2011nu} is the most general single-field inflation based on the Horndeski theory \cite{Horndeski:1974wa}, which is the most general single scalar-tensor theory with second-order field equations.
The Lagrangian has four arbitrary functions of a scalar field $\phi$ and $X\equiv -\partial_\mu\phi \partial^\mu \phi/2$:
\begin{equation}
\begin{split}
\mathcal{L}_2&=K(\phi,X),\\
\mathcal{L}_3&=-G_3(\phi,X)\Box\phi,\\
\mathcal{L}_4&=G_4(\phi,X)R+G_{4X}[(\Box\phi)^2-(\nabla_\mu \nabla_\nu \phi)^2],\\
\mathcal{L}_5&=G_5(\phi,X)G_{\mu\nu}\nabla^\mu \nabla^\nu \phi
-\frac{1}{6}G_{5X}[(\Box\phi)^3-3(\Box\phi)(\nabla_\mu \nabla_\nu \phi)^2+2(\nabla_\mu \nabla_\nu \phi)^3].
\end{split}
\end{equation}
where $R$ is the Ricci scalar and $G_{\mu\nu}$ is the Einstein tensor, $G_{iX}=\partial G_{i}/\partial X$, $(\nabla_\mu \nabla_\nu \phi)^2=\nabla_\mu \nabla_\nu \phi \nabla^\mu \nabla^\nu \phi$, and $(\nabla_\mu \nabla_\nu \phi)^3=\nabla_\mu \nabla_\nu \phi \nabla^\nu \nabla^\lambda \phi \nabla_\lambda \nabla^\mu \phi$.
We use the unitary gauge in which $\phi=\phi(t)$ and write the metric as
\begin{equation}
g_{00}=-1,
\qquad
g_{0i}=0,
\qquad
g_{ij}=a^2(t)(e^h)_{ij},
\end{equation}
where
\begin{equation}
(e^h)_{ij}=\delta_{ij} + h_{ij} + \frac{1}{2} h_{ik}h_{kj} + \frac{1}{6} h_{ik}h_{kl}h_{lj} + \cdots,
\end{equation}
to focus on the tensor perturbations.
Then the quadratic and cubic actions for $h_{ij}$ are obtained as below.
\begin{equation}
\begin{split}
S^{(2)}&=\frac{1}{8}\int dt \, d^3x \, a^3\left[\mathcal{G}_T \dot h^2_{ij}-\frac{\mathcal{F}_T}{a^2}(\partial_k h_{ij})^2\right],\\
S^{(3)}&=\int dt \, d^3x \, a^3\left[ \frac{\mathcal{F}_T}{4a^2}\left(h_{ik} h_{jl}-\frac{1}{2}h_{ij}h_{kl}\right)\partial_k \partial_l h_{ij} 
 + \frac{X\dot \phi G_{5X}}{12}\dot h_{ij}\dot h_{jk}\dot h_{ki} \right],
\end{split}
\end{equation}
where 
\begin{equation}
\begin{split}
\mathcal{F}_T&\equiv 2[G_4-X(\ddot\phi G_{5X}+G_{5\phi})],\\
\mathcal{G}_T&\equiv 2[G_4-2XG_{4X} -X(H\dot\phi G_{5X}-G_{5\phi})].
\end{split}
\end{equation}

From the quadratic action, the primordial power spectrum of gravitational waves, $\xi^{(s)}(\bfk)\equiv h_{jk}(\bfk) e^{*(s)}_{jk}(\bfk)$, of which $(s)$ stands for the helicity, is given by
\begin{equation}
\begin{split}
\langle \xi^{(s)}(\mathbf{k})\xi^{*(s')}(\mathbf{k}') \rangle &= (2\pi)^3 \delta^{(3)}(\mathbf{k}-\mathbf{k}') \delta_{ss'} \frac{\pi^2}{k^3} \mathcal{P}_h(k),\\
\mathcal{P}_{h}(k) &=  \frac{2H^2}{\pi^2}\frac{\mathcal{G}_T^{1/2} }{\mathcal{F}_T^{3/2}}.
\end{split}
\end{equation}
where $e^{(s_i)}_{jk}=e^{(s_i)}_{jk}(\bfk_i)$ is a transverse and traceless polarization tensor and it obeys the normalization condition $e^{(s)}_{jk}(\bfk)e^{*(s')}_{jk}(\bfk)=\delta_{ss'}$.
From the cubic action, we get the bispectrum of primordial gravitational waves with $\mathcal{P}_h^2$ \cite{Gao:2011vs}
\begin{equation}
\langle \xi^{(s_1)}(\bfk_1) \xi^{(s_2)}(\bfk_2) \xi^{(s_3)}(\bfk_3) \rangle
=(2 \pi)^7 \delta^{(3)}(\bfk_1 + \bfk_2 + \bfk_3) \: \mathcal{P}_h^2 \: \frac{\mathcal{S}^{s_1 s_2 s_3}_{\mathrm{(GR)}} + \mathcal{S}^{s_1 s_2 s_3}_{(5X)}}{k_1^2 k_2^2 k_3^2}, \label{eq:3pt}
\end{equation}
\begin{align}
&\mathcal{S}^{s_1 s_2 s_3}_{\mathrm{(GR)}}  = - \frac{1}{8} \frac{k_t}{k_1k_2k_3} \left( 1 - \frac{\sum_{i \neq j}k_i^2 k_j + 4\,k_1 k_2 k_3}{k_t^3} \right)  \left[ k_{3k} e^{*(s_1)}_{ki} e^{*(s_3)}_{ij} e^{*(s_2)}_{jl} k_{3l} 
+ \frac{1}{2} k_{2i} e^{*(s_1)}_{ik} k_{3k} e^{*(s_2)}_{jl} e^{*(s_3)}_{jl} \right]  \nonumber \\
&\qquad\qquad\qquad\qquad\qquad\qquad\qquad\qquad\qquad\qquad\qquad\qquad\qquad
\qquad
 + \text{2 circulations}, \label{eq:3ptgr}\\
&\mathcal{S}^{s_1 s_2 s_3}_{(5X)} = -\frac{1}{8}f_{5X}\frac{k_1 k_2 k_3}{k_t^3}
\ e^{*(s_1)}_{ij} e^{*(s_2)}_{jk} e^{*(s_3)}_{ki}, \label{eq:3ptnew}
\end{align}
where $k_t=k_1+k_2+k_3$, and $f_{5X}$ is given by
\begin{equation}
\begin{split}
&f_{5X} \equiv \frac{-2HX\dot \phi G_{5X}}{\mathcal{G}_T}.\\
\end{split}
\end{equation}

\section{Calculation of the bispectrum\label{calculation}}
Under linear evolution of the perturbations, the $a_{\ell m}$'s are given as
\begin{equation}
a^{(s)}_{\ell m} = 4\pi (-i)^\ell \int \frac{k^2 dk}{(2\pi)^3} \mathcal{T}^{(s)}_\ell(k) \int d\Omega_{k}\; {}_{-s}Y^*_{\ell m} \xi^{(s)}(\bfk),
\end{equation}
where the superscript $(s)$ stands for the helicity of cosmological perturbations which induce the CMB anisotropies.
Then the bispectrum is represented as the sum of all the helicity contributions
\begin{equation}
B_{\ell_1 \ell_2 \ell_3} = \sum_{s_1 s_2 s_3} B^{(s_1 s_2 s_3)}_{\ell_1 \ell_2 \ell_3},
\end{equation}
\begin{equation}
\begin{split}
B^{(s_1 s_2 s_3)}_{\ell_1 \ell_2 \ell_3}
&=\sum_{m_1 m_2 m_3} \wtj{\ell_1}{\ell_2}{\ell_3}{m_1}{m_2}{m_3}\\
 &~~~~\times \prod_{j=1}^{3} \left[ 4\pi (-i)^{\ell_j} \int \frac{k_j^2 dk_j}{(2\pi)^3} \mathcal{T}^{(s_j)}_{\ell_j}(k_j) 
 \int d\Omega_{k_j} {}_{-s_j}Y^*_{\ell_j m_j} \right]
 \langle \xi^{(s_1)}(\bfk_1) \xi^{(s_2)}(\bfk_2) \xi^{(s_3)}(\bfk_3) \rangle.
\end{split}
\end{equation}

Since the Horndeski theory does not violate the parity symmetry, the bispectum vanishes when the total multipole moment is even or odd, depending on the parity of the types of the CMB fluctuations.
For example, the auto-bispectrum of B-mode polarization is
\begin{equation}
B_{\ell_1 \ell_2 \ell_3}
=0 \quad \mathrm{for} \quad \ell_1+\ell_2 +\ell_3=\mathrm{even},
\end{equation}
and the auto-bispectrum of E-mode polarization is
\begin{equation}
B_{\ell_1 \ell_2 \ell_3}
=0 \quad \mathrm{for} \quad \ell_1+\ell_2 +\ell_3=\mathrm{odd}.
\end{equation}

Replacing the polarization tensors with the spin-weighted spherical harmonics, we get the bispectrum originated from the non-Gaussianities of primordial gravitaional waves with the Wigner-$9j$ symbol,
\begin{align}\label{eq:integration}
B^{(s_1 s_2 s_3)}_{\ell_1 \ell_2 \ell_3}
&=\sum_{\substack{\ell'_1\ell'_2\ell'_3\\L_1L_2L_3}} 
\mathcal{I}_{\ell_1\ \ell'_1\ L_1}^{s_1-s_1\, 0}
\mathcal{I}_{\ell_2\ \ell'_2\ L_2}^{s_2-s_2\, 0}
\mathcal{I}_{\ell_3\ \ell'_3\ L_3}^{s_3-s_3\, 0}
\mathcal{I}_{L_1L_2L_3}^{0\ 0\ 0} 
\wnj{\ell_1}{\ell_2}{\ell_3}{\ell'_1}{\ell'_2}{\ell'_3}{L_1}{L_2}{L_3}
\nonumber\\
&\times \int x^2dx \ \prod_{j=1}^{3} \left[ \frac{2}{\pi}\  i^{L_j-\ell_j} \int dk_j \mathcal{T}^{(s_j)}_{\ell_j}(k_j) j_{L_j}(k_j x) 
\right] \nonumber\\
&\times (2\pi)^4 \mathcal{P}_h^2 \; \mathcal{K}(k_1,k_2,k_3),
\end{align}
where $\mathcal{K}=\mathcal{K}_{(\textrm{GR})}+\mathcal{K}_{(5X)}$ represents the wavenumber dependence of the non-Gaussianities with
\begin{align}
&\mathcal{K}_{(\textrm{GR})}=  \frac{(4\pi)^{3/2}}{80} \left( k_t - \frac{\sum_{i \neq j}k_i^2 k_j + 4\,k_1 k_2 k_3}{k_t^2} \right) \left[ \frac{k_3}{k_1k_2} \delta_{\ell'_1\,2}\,\delta_{\ell'_2\,2}\left( \frac{ 5\delta_{\ell'_3\,2}}{\sqrt{21}}-\frac{\delta_{\ell'_3\,4}}{\sqrt{7}} \right) 
+ \textrm{2 circulations}\right],\\
&\mathcal{K}_{(5X)}=  \frac{(4\pi)^{3/2}}{80} \; f_{5X} \sqrt{\frac{7}{3}}  \frac{k_1k_2k_3}{k_t^3} \delta_{\ell'_1\,2}\,\delta_{\ell'_2\,2}\,\delta_{\ell'_3\,2}.
\end{align}

Due to the complicated form of $\mathcal{K}(k_1,k_2,k_3)$, $k_j$ integration in \eqref{eq:integration} cannot be performed separately.
By using the Laplace transformation,
\begin{equation}
p^{-\nu-1}=[\Gamma(\nu+1)]^{-1} \int_0^\infty t^\nu e^{-pt} dt \qquad (\mathrm{Re}[\nu]>-1), 
\end{equation}
however, $k_t^{-\nu-1}$ becomes a product of functions of $k_j$'s, to make each term in \eqref{eq:integration} a product of functions of $k_1$, $k_2$ and $k_3$.
Although a new integration variable must be introduced here, the exponential factor decays so rapidly that the computation cost can be reduced as in separable templates of non-Gaussianities \cite{Smith:2006ud}. 
We emphasize that the laplace transformation enables us to get the exact primary bispectra of B-mode polarization up to the nonlinearity parameter $f_{5X}$, 
in contrast to the temperature bispectrum from primordial scalar perturbations which have been studied based on a few templates that are not necessarily derived from specific models of inflation \cite{Komatsu:2010hc,Yadav:2010fz}.

We calculated evolution of linear perturbations with the Boltzmann code CLASS \cite{Blas:2011rf} with the cosmological parameters estimated by the latest observation \cite{Ade:2015xua} to obtain the transfer functions. 
We assume the spectrum of primordial gravitational waves is scale invariant, and their amplitude is parametrized by the tensor-to-scalar ratio $r$.

\section{Results\label{results}}
\begin{figure}[htbp]
\includegraphics[width=7cm]{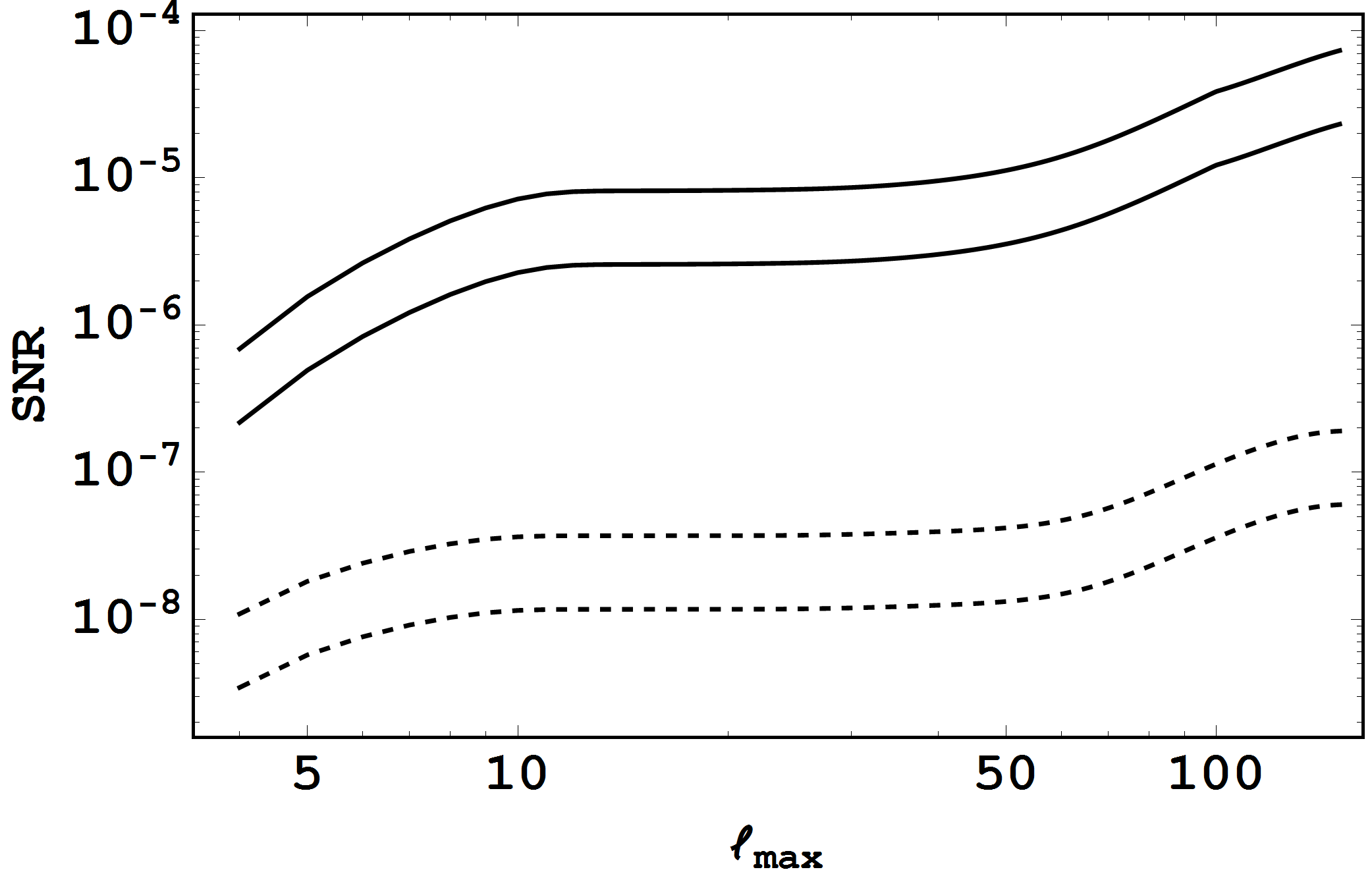}
\caption{
Signal-to-noise ratio when B-mode polarization is observed up to the maximum of multipole moment $\ell_\mathrm{max}$.
The bold lines are for general relativity and the dotted lines are for the new class of gravity whose parameter $f_{5X}=1$.
The upper lines are for tensor-to-scalar ratio $r=0.1$, and the lower lines are for $r=0.01$. 
\label{fig:SNRbbb}}
\end{figure}

Once the B-mode auto-bispectrum produced by primordial gravitational non-Gaussianity has been calculated,
the maximum possible signal-to-noise ratio where observational errors are dominated by the cosmic variance is given by $(S/N)=1/\sqrt{F^{-1}}$, where $F$ is the Fisher information defined by
\begin{align}
F=\sum_{\ell_1,\ell_2,\ell_3\le\ell_\text{max}} \frac{B^2_{\ell_1 \ell_2 \ell_3}}{6\, C_{\ell_1}C_{\ell_2}C_{\ell_3}}.
\end{align}
First, we supposed that $f_{5X}=0$ and found that the signal is too small to be detected.
We show the signal-to-noise ratios in the zero-noise ideal experiment as bold lines in the figure \ref{fig:SNRbbb}.
Next we assume the case in which $f_{5X}$ is large enough for its non-Gaussianity to dominate.
If the coefficient $f_{5X}$ is as large as $10^7$ or $3\times10^7$, we are able to detect the bispectrum when tensor-to-scalar ratio $r=0.1$ or $0.01$, respectively, by an observation up to $\ell_\mathrm{max}=200$.

We also calculated CMB temperature bispectrum induced by this mode and found that even if $f_{5X}$ takes such a large value, the signal-to-noise ratio of the resultant temperature bispectrum is too small to be detected, so that it would not contradict any current observational data.

\begin{figure*}[htbp] 
  \begin{center}
      \subfloat[Normalized $\left|\mathcal{S}^{s_1 s_2 s_3}_{\mathrm{(GR)}}\right|$.]{
       \includegraphics[clip,width=7cm]{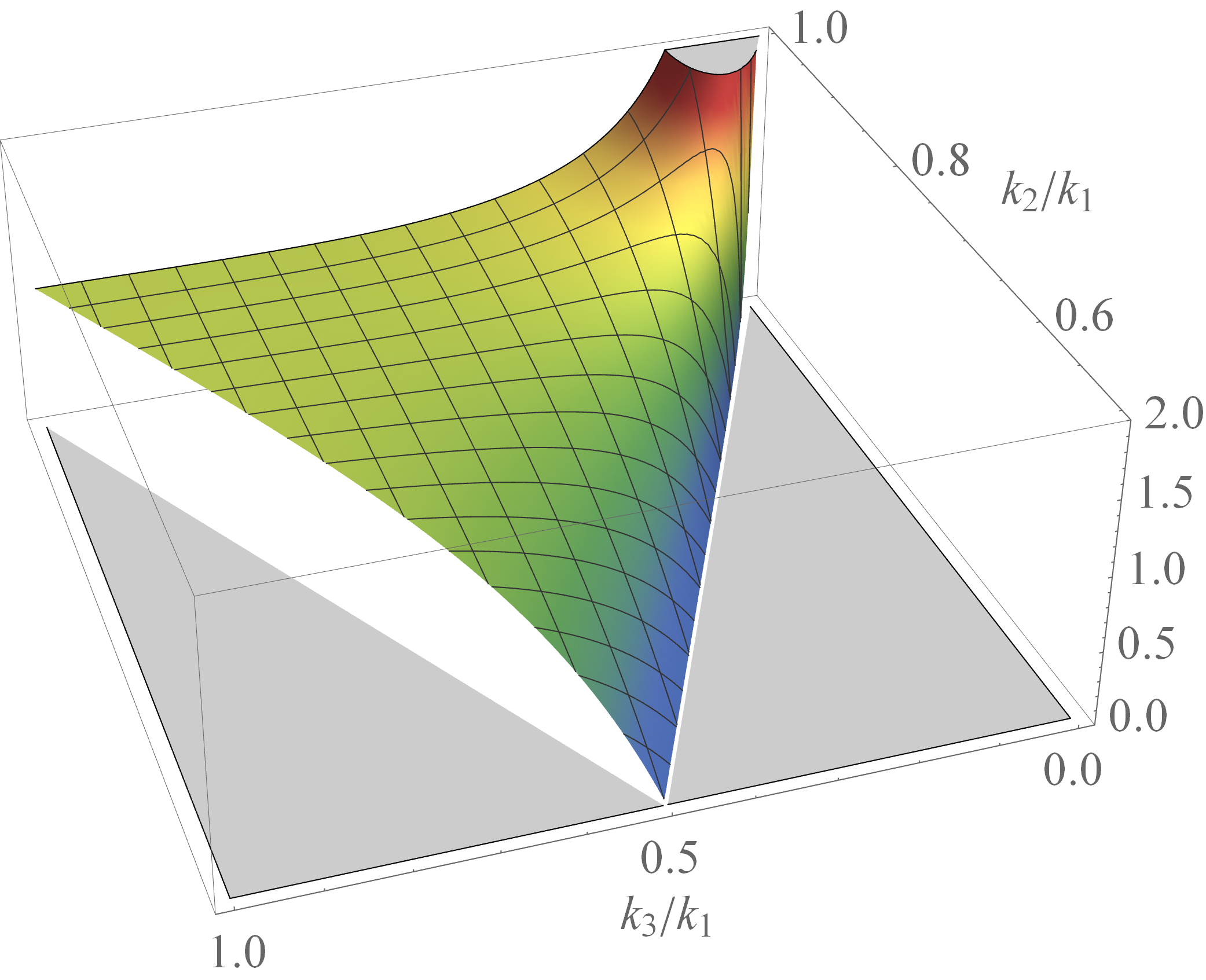}
      }
      \subfloat[Normalized $\left|\mathcal{S}^{s_1 s_2 s_3}_{(5X)}\right|$.]{
       \includegraphics[clip,width=7cm]{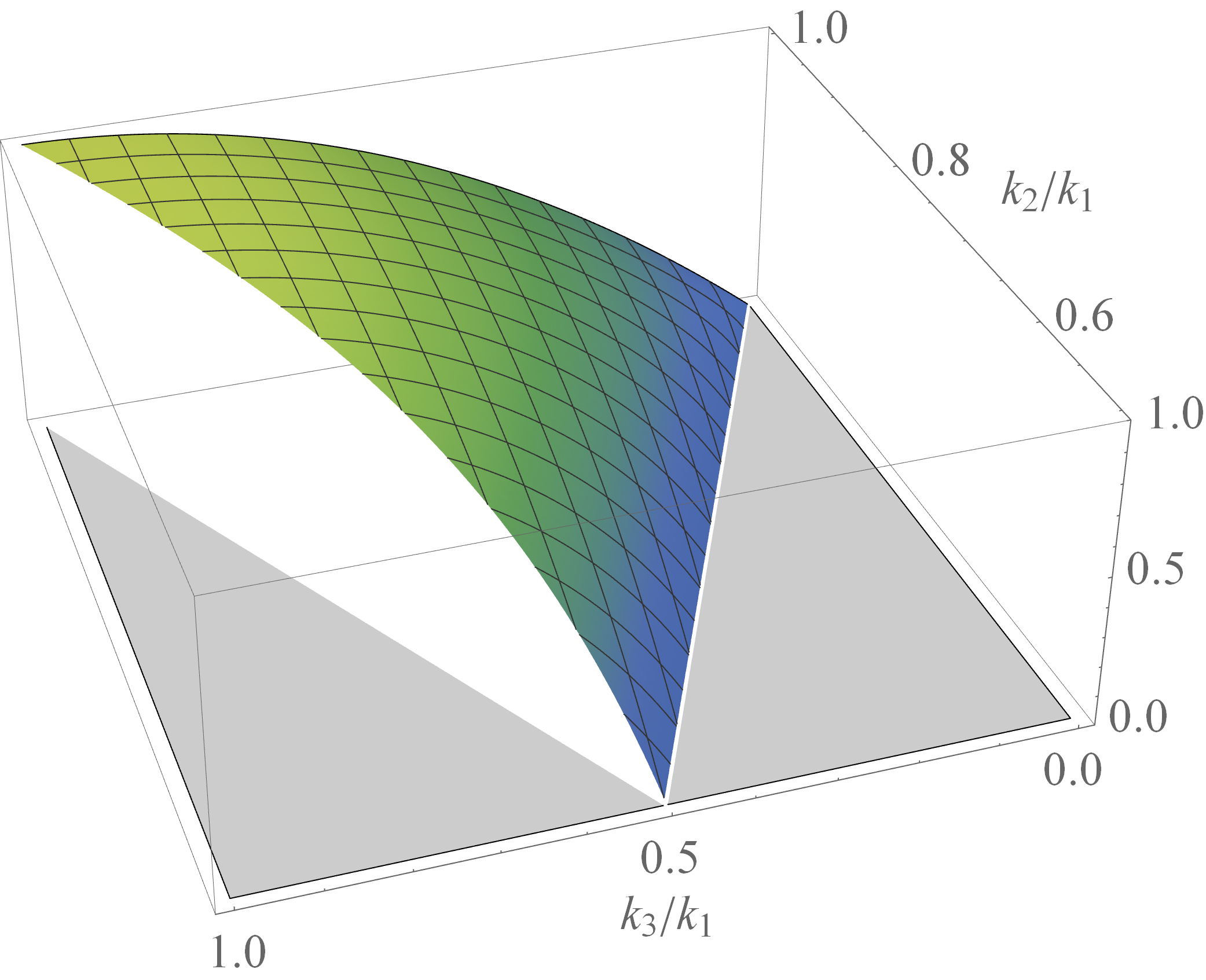}
      }
    \caption{
      The shape function $\left|\mathcal{S}^{s_1 s_2 s_3}_{\mathrm{(GR}/5X)}\right|$ 
      of non-Gaussianity of primordial gravitational waves from (a) general relativity and (b) nonzero $G_{5X}$.
      The vertical axis is normalized to unity for $k_2/k_1=k_3/k_1=1$. \label{fig:shape}
    }
  \end{center}
\end{figure*}

\begin{figure*}[htbp]
  \begin{center}
      \subfloat[$B_{\ell_1 \ell_2 \ell_3}$ induced by general relativity.]{
        \includegraphics[clip,width=7cm]{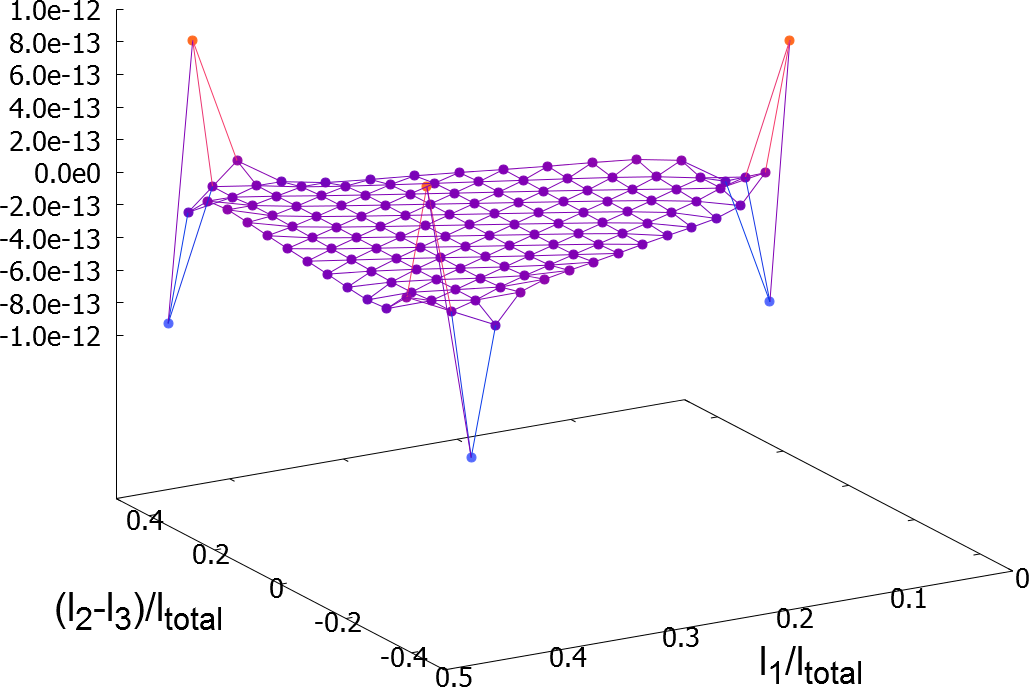}
      }
      \subfloat[$B_{\ell_1 \ell_2 \ell_3}$ induced by nonzero $G_{5X}$.]{
        \includegraphics[clip,width=7cm]{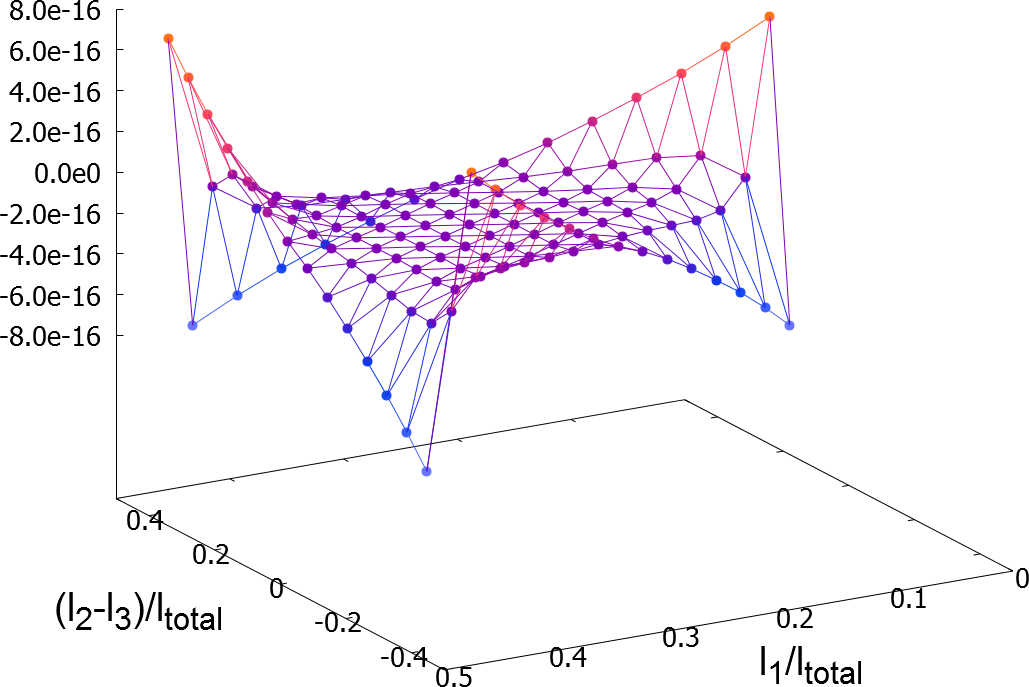}
      }
    \caption{
        The B-mode bispectrum $B_{\ell_1 \ell_2 \ell_3}$ induced by (a) general relativity
        and (b) nonzero $G_{5X}$ where $\ell_1+\ell_2+\ell_3=33$. 
        Here we divide $B_{\ell_1 \ell_2 \ell_3}$ by (a) $i \mathcal{P}_h^2$ 
        and (b) $i \mathcal{P}_h^2 f_{5X}$ respectively.  \label{fig:shapeBBB}
    }
  \end{center}
\end{figure*}

We show the shapes of the primordial non-Gaussianities and the B-mode reduced bispectrum
(see figure \ref{fig:shape}-\ref{fig:shapeBBB}.)
Their shapes are similar in both general relativity and the new class of gravity in which $G_{5X}$ does not vanish.

\begin{figure}
\includegraphics[width=5cm]{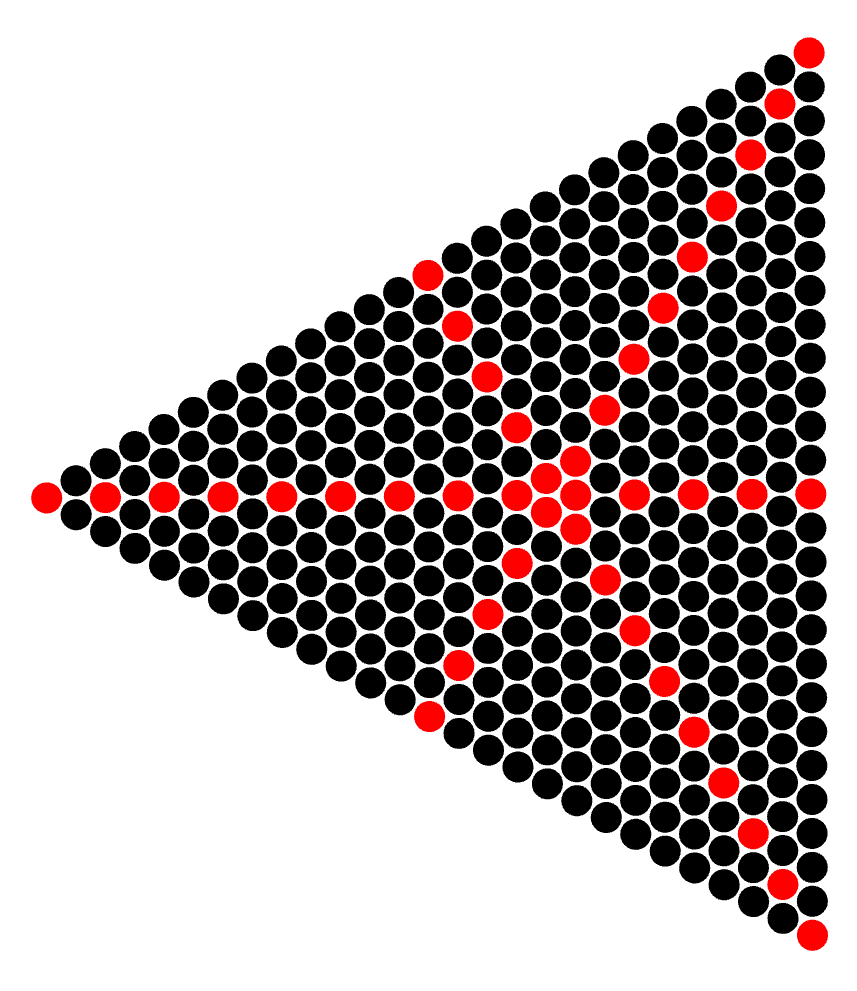}
\caption{An example of points where at least two of $\ell_j$ are equal in multipole moment space $(\ell_1,\ell_2,\ell_3)$ where $\ell_1+\ell_2+\ell_3=\mathrm{const.}$\label{fig:parityCond}}
\end{figure}

The B-mode bispectrum $B_{\ell_1 \ell_2 \ell_3}$ is zero along the lines (see figure \ref{fig:parityCond}) where at least two of $\ell_j$ are equal, since Wigner-$3j$ symbol $\wtj{\ell_1}{\ell_2}{\ell_3}{m_1}{m_2}{m_3}$ is antisymmetric to a replacement of columns when $\ell_1+\ell_2 +\ell_3=\mathrm{odd}$.
This nature comes from the odd parity of B-mode bispectrum.
The bispectrum decreases toward the lines and so the signal-to-noise ratio is smaller than that of the E-mode auto-bispectrum which is not suppressed because it has the even parity.

\section{conclusion\label{conclusions}}
In the paper, we calculated the B-mode auto-bispectrum induced by tensor perturbations produced during inflation in the early Universe based on the most general single field inflation model \cite{Gao:2011vs}.
In this theory it is known that the primordial bispectrum of tensor perturbations consist of two distinct terms, one present in general relativity, and the other propotional to $f_{5X}$.
We have found that the former contribution, whose nonlinearity parameter is always unity, cannot be detected by the CMB, whereas the new term may be detectable only if $f_{5X}\gtrsim10^7$

As a future prospect, when 21-cm maps of atomic hydrogen distribution in the dark ages become available, the curl mode of gravitational lensing effect \cite{Book:2011dz} may be the key of detection of the non-Gaussianity of primordial gravitational waves, because 21-cm fluctuations be seen up to $\ell\sim 10^7$ in principle and there are many statistically independent fluctuations.

\begin{acknowledgments}
We are grateful to Eiichiro Komatsu for a useful comment in the beginning of the work.
HWHT was supported by the Advanced Leading Graduate Course for Photon Science (ALPS).
This work was supported by JSPS KAKENHI, Grant-in-Aid for Scientific Research 15H02082 and
Grant-in-Aid for Scientific Research on Innovative Areas 15H05888.
\end{acknowledgments}

\bibliography{mypaper}

\end{document}